\documentclass[10.5pt]{article}
 
\usepackage[margin=0.75in]{geometry} 
\usepackage{amsmath,amsthm,amssymb,scrextend,parskip}
\usepackage{dsfont}

\usepackage{fancyhdr}
\usepackage[sort&compress]{natbib}
\usepackage{graphicx}
\usepackage{enumerate}
\usepackage{bbm}
\usepackage{mathtools}
\usepackage{xcolor} 
\usepackage{url}
\usepackage{authblk}

\usepackage{scalerel}

\setcitestyle{numbers,square}

\newcommand{\Z}{\mathbb{Z}}

\newcommand{\cS}{\mathcal{S}}
\newcommand{\cH}{\mathcal{H}}

\newcommand{\sth}{ \ \mathrm{s.t.} \ }
\newcommand{\eps}{ \epsilon }
\newcommand{\id}{\mathbbm{1}}
\newcommand{\cptp}{\mathcal{E}}
\newcommand{\rank}{\operatorname{rank}}
\newcommand*{\defeq}{=}                     
                     
\DeclarePairedDelimiter\ceil{\lceil}{\rceil}
\DeclarePairedDelimiter\floor{\lfloor}{\rfloor}
                     
\newcommand{\pmac}{\prec}
\newcommand{\maj}{\prec_{\mathrm{M}}}
\newcommand{\pmaj}{\prec_\mathrm{p}}

\newcommand{\EqSt}{\Gamma_{\mathrm{EQ}}}

\newcommand{\tr}[1]{\operatorname{tr}\left( #1 \right)}
\newcommand{\partr}[2]{\operatorname{tr}_{#2}\left( #1 \right)}

\newcommand{\epsball}[2]{\mathcal{B}^{#1}\left( #2 \right)}
\newcommand{\epsballsub}[2]{\mathcal{B}^{#1}_\leq \left( #2 \right)}
\newcommand{\trdist}[2]{\operatorname{D}\left( #1 , #2 \right)}

\newcommand{\Hmin}[1]{H_{\min}\left( #1 \right)}
\newcommand{\Hzero}[1]{H_{0}\left( #1 \right)}

\newcommand{\Hmineps}[2]{H^{#1}_{\min}\left( #2 \right)}

\newcommand{\HminepsO}[1]{H^{#1}_{\min}}
\newcommand{\HHeps}[2]{H^{#1}_{\mathrm{H}}\left( #2 \right)}
\newcommand{\HHepsO}[1]{H^{#1}_{\mathrm{H}}}

\newcommand{\maxEV}{\lambda_{\max}}

\newcommand{\ithEV}[1]{\lambda_{{i}}\left( #1 \right)}
\newcommand{\ithEVk}[2]{\lambda_{{#2}}\left( #1 \right)}

\newcommand{\ketbra}[2]{|#1\rangle\! \langle#2|}

\newcommand{\refSt}{\rho_\mathrm{ref}}
\newcommand{\refStph}{X_\mathrm{ref}}

\renewcommand{\emph}[1]{\textit{#1}}

\newtheoremstyle{theorem}
	{6pt}
	{}
	{\itshape}
	{}
	{\bfseries}
	{:}
	{.5em}
	{}
\theoremstyle{theorem}
\newtheorem{thm}{Theorem}[section]

\newtheorem{ex}{Example}[section]
\newtheorem{cor}{Corollary}[section]
\newtheorem{prop}{Proposition}[section]

\newtheoremstyle{defn}
	{6pt}
	{}
	{}
	{}
	{\bfseries}
	{:}
	{.5em}
	{}
\theoremstyle{defn}
\newtheorem{defn}[]{Definition}[section]

\makeatletter
\colorlet{phfcolor}{red!50!orange}
\colorlet{phfrmcolor}{phfcolor!50!gray!35!white}
\colorlet{phfrmcolorlink}{blue!40!phfrmcolor}
\def\phf{\@ifnextchar[\phf@c\phf@}
\def\phf@{\@ifstar\phf@s\phf@t}
\newcommand\phf@c[1][]{{\color{phfcolor}\fontfamily{\sfdefault}\fontseries{sb}\selectfont \;[\,{#1}\,]\;}}
\newcommand\phf@t[1]{{\color{phfcolor}{#1}}}
\newcommand\phf@s[1]{{\colorlet{docnotelinkcolor}{phfrmcolorlink}\color{phfrmcolor}{\itshape #1}}}
\robustify\phf
\makeatother

\newcommand{\tit}{Smooth entropy in axiomatic thermodynamics}

\begin{document}
\lhead{\tit}

\rhead{ \today}

\title{\tit}
\author[1]{Mirjam Weilenmann}
\affil[1]{Department of Mathematics, University of York, Heslington, York, YO10 5DD, UK}
\author[2]{Lea Kr{\"a}mer Gabriel}
\affil[2]{Institute for Theoretical Physics, ETH Zurich, 8093 Switzerland}
\author[3]{Philippe Faist}
\affil[3]{Institute for Quantum Information and Matter, California Institute of Technology, Pasadena 91125, USA}
\author[2]{Renato Renner}

\date{{\small \today}}

\setcounter{Maxaffil}{0}
\renewcommand\Affilfont{\itshape\small}

\makeatletter
\renewcommand\AB@authnote[1]{\rlap{\textsuperscript{\normalfont#1}}}
\renewcommand\Authsep{,~\,}
\renewcommand\Authands{,~\,and }
\makeatother

\maketitle

\begin{abstract}
Thermodynamics can be formulated in either of two approaches, the phenomenological approach, which refers to the macroscopic properties of systems, and the statistical approach, which describes systems in terms of their   microscopic constituents. We establish a connection between these two approaches by means of a new axiomatic framework that can take errors and imprecisions into account. This link extends to systems of arbitrary sizes including microscopic systems, for which the treatment of imprecisions is pertinent to any realistic situation.
Based on this, we identify the quantities that characterise whether certain thermodynamic processes are possible with entropy measures from information theory. In the error-tolerant case, these entropies are so-called smooth min and max entropies. Our considerations further show that in an appropriate macroscopic limit there is a single entropy measure that characterises which state transformations are possible. In the case of many independent copies of a system (the so-called i.i.d.\ regime), the relevant quantity is the von Neumann entropy. 
\end{abstract}


\section{Introduction}
The thermodynamic behaviour of macroscopic systems is traditionally described
according to either of two theories: we can take a phenomenological approach
that refers to macroscopic quantities such as the volume and the pressure of a
system in which the possible thermodynamic processes are constrained via
  the traditional laws of thermodynamics, or we can take a statistical approach that
  begins with a microscopic description of the system, for instance by
  describing the motion of individual particles, and then infers the system's
  corresponding macroscopic properties via statistical and typicality
  considerations.  The two theories are fundamentally different, each referring
to quantities that are not defined within the other, but at the same time they
are known to lead to consistent descriptions of the
behaviour of thermodynamic systems in equilibrium. The two approaches are usually related by
  studying specific examples and connecting the corresponding physical
  quantities, such as identifying the Boltzmann entropy with the thermodynamic
  entropy in the case of an isolated ideal gas or a spin chain.  In standard
  textbooks, such identifications are largely carried out based on the
  properties these quantities display: For instance, the Boltzmann entropy and
  the thermodynamic entropy are both extensive, and their derivatives with
  respect to the total energy are both equal to the inverse temperature.  While
  the correspondence between these approaches clearly holds at a very universal
  level, a precise and general connection is hindered by the very different
  underlying physical frameworks assumed in either approach.

The aim of this chapter is to describe the phenomenological and the statical approach on the same footing, i.e., using the same framework. Rather than
focusing on derived quantities such
as the entropy, we connect the two approaches on the level of their basic
underlying structure. This structure is rooted in identifying the
  possible processes that can occur spontaneously.  While it is rather naturally suited to phenomenological thermodynamics, and can be interpreted
  as an abstract version thereof, it also applies to the
  microscopic realm pertinent to the statistical approach.  In
  this abstract structure, whether or not a process may occur is characterized
  by a specific quantity---which is called, by extension, an \emph{entropy
    function}.  In the phenomenological approach, it coincides with the
  thermodynamic entropy and naturally characterises the possible adiabatic
  processes. 

It turns out that, in order to relate the phenomenological to the statistical approach, it is critical to take small perturbations into account. One must allow a degree of imprecision in processes in order to obtain physical results that do not depend on unobservable features of the quantum state, such
  as distribution tails (for instance, when isothermally compressing a gas, one
  typically ignores the overwhelmingly unlikely event in which all particles
  conspire to hit against the piston, which would require more invested work).
  We show that the abstract structure of possible processes extends naturally to
  the case where imprecisions are to be taken into account.  In an
  appropriate macroscopic limit, such as by considering a large system composed
  of many independent particles, we recover usual phenomenological thermodynamics.  That is, we identify a class of macroscopic states
  whose structure of possible processes coincides with that of the
  usual phenomenological approach.  Hence, beyond the fact that both the
  phenomenological approach and the statistical approach can be phrased within
  the same type of framework, the former can be obtained from the latter on an
  abstract level, thus drawing a robust and general connection between them that transcends
  individual examples.

To establish our results, we rely on a resource theoretic formulation of
thermodynamics. There are several approaches to this, depending on which
operations are considered to be the \emph{free operations} of the resource
theory.  The most widespread approach relies on thermal
operations~\cite{Janzing2000_cost,Horodecki2011,Renes2014}, which are reviewed
in~\cite{Ng2018arXiv_TO}.  Instead, we consider
\emph{adiabatic processes} here, which are central to Lieb and Yngvason's
axiomatic approach to phenomenological
thermodynamics~\cite{Lieb1998,Lieb1999,Lieb2001,Lieb2013,Lieb2014,Thess}. Their
approach follows a long tradition of formulating thermodynamics
axiomatically~\cite{Caratheodory1909,%
  Giles1964,%
  Lieb1998,%
  Lieb1999,%
  Lieb2001,%
  Lieb2013,%
  Lieb2014,%
  Thess,%
  gyftopoulos1991thermodynamics,%
  Gyftopoulos2005,%
  Zanchini2011_thermodynamics,%
  Zanchini2014,%
  Hatsopoulos1,%
  Hatsopoulos2a,%
  Hatsopoulos2b,%
  Hatsopoulos3%
}, which has also been continued by recent work~\cite{LidiaPhD, LeaPhD,
  Hulse2018, Kammerlander2018}.  In an adiabatic process, a system interacts
with a weight that can perform or extract work from the system without changing
the environment. Contrary to the thermal operations, these processes do not
impose any constraint regarding the conservation of energy. Instead, they forbid
the equilibration with a reservoir, which would result in a change of the state
of the system's environment. We introduce Lieb and Yngvason's
axiomatic framework for thermodynamics in Section~\ref{sec:resource} and relate the notion of an adiabatic process to the statistical picture in
Section~\ref{sec:noisy}. Based on this, we present an axiomatic
relation between the entropy measures relevant for thermodynamics in the
phenomenological and in the statistical approach. This is essentially a summary of results derived in Ref.~\cite{Weilenmann2015}.

Furthermore, we introduce a novel, error-tolerant axiomatic framework that allows us to realistically describe systems at any scale, including microscopic and macroscopic systems. (Previous approaches were typically concerned solely with macroscopic systems and did not take errors into account.)  
Our framework contributes to the current development of pushing resource theories towards more realistic regimes, which has been initiated through work on probabilistic transformations~\cite{Alhambra2015} and finite size effects~\cite{Gemmer2006, Sparaciari2016} and which has recently also led to the study of imprecisions in specific resource theories~\cite{VanderMeer2017, Horodecki2017, Hanson2017, Hanson2017a, thesis}. 
Similar to our consideration of Lieb and Yngvason's work, we present our new framework first phenomenologically in Section~\ref{sec:error} and then from the statistical viewpoint in Section~\ref{sec:microerror}. This allows us to  relate the quantities that characterise whether there exists an error-tolerant adiabatic process between certain states to smooth min and max entropies, which are known from single-shot information theory~\cite{PhdRenner2005_SQKD,TomamichelBook,Dupuis2013_DH}.

In the limit of large systems, our error-tolerant framework furthermore recovers the structure of a resource theory for macroscopic thermodynamics. Specifically,  there is a single entropy function that specifies whether a transformation between different equilibrium states of a macroscopic system is possible. 
In the spirit of the chapter, we present these results first according to the phenomenological approach in Section~\ref{sec:emergence}, followed by the statistical perspective in Section~\ref{sec:examples}. The latter viewpoint allows us to recover the von Neumann entropy as the quantity that characterises  the behaviour of so-called i.i.d.\ states under error-tolerant adiabatic processes. 
Through this consideration of macroscopic systems in equilibrium we relate our error-tolerant framework to the framework for thermodynamics introduced by Lieb and Yngvason. Our elaborations in Sections~\ref{sec:error} to \ref{sec:examples} regarding the error-tolerant axiomatic  framework and its macroscopic limits are based on~\cite{thesis}.

\section{Axiomatic framework for phenomenological thermodynamics}\label{sec:resource}

In phenomenological thermodynamics, the state of a system in equilibrium is usually described in terms of a few real and positive parameters. For instance, the state of a gas in a box is often given in terms of its internal energy, $U$, and its volume, $V$. Such an equilibrium state $X=(U,V)$ lives in the space of all   equilibrium states of that system, denoted $\EqSt$. The description of the state of a system out of equilibrium is more involved; such states live in a larger state space $\Gamma \supseteq \EqSt$.
 
Thermodynamics is a resource theory, a perspective that is implicit in the axiomatic framework proposed by Lieb and Yngvason to derive the second law~\cite{Lieb1998,Lieb1999,Lieb2001,Lieb2013,Lieb2014}.
At the core of this framework lie the \emph{adiabatic processes}. They are  defined in~\cite{Lieb1998,Lieb1999,Lieb2001,Lieb2013,Lieb2014} as those operations on a system that leave no trace on its environment, except for a change in the position of a weight. These processes are sometimes also called \emph{work processes}~\cite{Gyftopoulos2005}. They induce a preorder relation $\pmac$ on the state space, $\Gamma$, of a system that is called \emph{adiabatic accessibility}.  For two states $X$, $Y \in \Gamma$, there is an adiabatic process transforming $X$ into $Y$ if and only if $X \pmac Y$.\footnote{
Throughout this chapter, we follow Lieb and Yngvason's convention regarding the notation of the order relation, where $X \pmac Y$ means that $X$ can be transformed into $Y$ with an adiabatic process. Note that this is the reverse convention of what a reader acquainted with the literature on quantum resource theories might expect. \label{fnt:notation}} States $X$, $Y \in \Gamma$ that can be adiabatically interconverted, i.e., $X \pmac Y$ as well as $Y \pmac X$, are denoted $X \sim Y$.

\begin{ex}\label{ex:adiab}
Let there be one mole of Helium gas in a box that is in a thermodynamic equilibrium state (which we may treat as an ideal monoatomic gas). Its state can be characterised by a tuple $(U,~V)$, where $U \geq 0$ is the internal energy and $V \geq 0$ is the volume of the gas. Now assume that there are two states of the gas that differ in their internal energy, for instance $X=(U,~V)$ and $Y=(2U,~V)$. Then we can find an adiabatic process transforming the state $X$ into $Y$, namely we can use a weight to mechanically stir the gas up to increase its internal energy, hence $X \pmac Y$. An adiabatic process that recovers $X$ again from $Y$ cannot be constructed. If we were to consider the state $Z=(2U,~V2^{-\frac{3}{2}})$ instead of $Y$, however, we could construct adiabatic processes that allow us to interconvert the two, i.e., $X \sim Z$. Namely, we could compress the gas adiabatically to obtain the state $Z$ (using a weight to quickly move a piston). To recover $X$, the gas could be decompressed changing nothing else in the environment than the position of a weight. This coincides with the textbook understanding of an adiabatic compression and decompression.
\end{ex}

The state of a composed system is denoted as the Cartesian product of the individual states of the systems to be composed, for $X \in \Gamma$ and $X' \in \Gamma'$ it is $(X,~X') \in \Gamma \times \Gamma'$.\footnote{The composition operation is assumed to be associative and commutative.} Physically, composition is understood as bringing individual systems together to one without letting them interact (yet). The state of the composed system is therefore naturally characterised by the properties of its constituent systems.
 
The set of equilibrium states of a thermodynamic system, $\EqSt$, is central to the axiomatic framework and will be necessary for introducing an entropy function later as well as for deriving a second law. In the axiomatic framework, equilibrium states are distinct from other states in that they are assumed to be scalable, i.e., for any $\alpha \in \mathbbm{R}_{\geq 0}$ one can define a scaled version of an equilibrium state $X \in \EqSt$, denoted as $\alpha X \in \alpha \EqSt$.\footnote{The scaling is required to obey $1 X = X$ as well as $\alpha_1(\alpha_2 X) = (\alpha_1 \alpha_2) X$, thus $1 \EqSt = \EqSt$ and $\alpha_1(\alpha_2 \EqSt) = (\alpha_1 \alpha_2) \EqSt$.}
Scaling a system by a factor $\alpha$ means taking $\alpha$ times the amount of substance of the original system. As is usual in phenomenological thermodynamics, we ignore the fact that the system is made up of a finite number of particles that cannot be subdivided, and allow the scaling by any real number.

\begin{ex}
Let us consider two systems, one being a mole of Helium gas in an equilibrium state $X=(U,~V)$, and the other two moles of Helium in another equilibrium state $X'=(U',~V')$. The composition operation allows us to regard these as subsystems of a composed system in state $(X,~X')$. Adiabatic operations may then either affect one of the individual subsystems or both of them. For instance, we may connect the two systems and allow them to equilibrate thermally by means of such an operation. Alternatively, removing the walls that are separating the two subsystems is an adiabatic process on the composite system, which in this case would lead to a system containing three moles of Helium gas.

Scaling the state $X$, say by a factor of $3$, leads to a state $3X=(3U,~3V)$, where the extensive properties of the state scale with the system size. Note that if the gas were not in an equilibrium state, for instance if it had a temperature gradient, it would not be clear how its properties scale with the amount of substance and the scaling operation would not be defined.
\end{ex}

For the order relation of adiabatic accessibility a few physically motivated properties shall be assumed~\cite{Lieb1998,Lieb1999,Lieb2001}. For equilibrium states $X,Y,Z \in \EqSt$ and $X',Y',Z_0,Z_1 \in \EqSt'$, these are:
\begin{enumerate}[(E1)]
\item Reflexivity: $X \sim X$.
\item Transitivity: $X \pmac Y$ and $Y \pmac Z$ $\implies$ $X \pmac Z$.
\item Consistent composition: $X \pmac Y$ and $X' \pmac Y'$ $\implies$ $(X,~X') \pmac (Y,~Y')$.
\item Scaling invariance: $X \pmac Y$ $\implies$ $\alpha X \pmac \alpha Y$, $\forall \ \alpha > 0$. 
\item Splitting and recombination: For $0<\alpha<1$, $X \sim (\alpha X,~(1-\alpha) X)$.
\item Stability: If $(X,~\alpha Z_{\mathrm{0}}) \pmac (Y,~\alpha Z_{\mathrm{1}})$ for a sequence of scaling factors $\alpha \in \mathbbm{R}$ tending to zero, 
then $X \pmac Y$.
\end{enumerate}
\begin{enumerate}[(CH)]
\item Comparison Hypothesis: For each $0 \leq \alpha \leq 1$, any two states in $(1-\alpha)\EqSt \times \alpha \EqSt$ can be related by means of $\prec$, i.e., for any $X, Y \in (1-\alpha)\EqSt \times \alpha \EqSt$ either $X \prec Y$ or $Y \prec X$ (or both).
\end{enumerate}

Axioms (E1)--(E4) are naturally obeyed by any order relation $\prec$ that is specified through a class of processes that can be composed sequentially as well as in parallel (on composed systems) and that respect a scaling operation. Notice that (E5) relates states on different spaces. In the example of a box of gas in an equilibrium state $X$ an adiabatic process interconverting the two could be realised by inserting and removing a partition at a ratio $\alpha : (1-\alpha)$. (E6) captures the physical intuition that arbitrarily small impurities in a large thermodynamic system should not affect the possible adiabatic processes.

Note that in~\cite{Lieb1998,Lieb1999,Lieb2001}, (CH) is not stated as an axiom but rather derived from additional axioms about thermodynamic systems. For this derivation more structure is required, such as the definition of so-called simple systems, the state of which is an $n$-tuple with one distinguished energy coordinate. The discussion of these additional axioms for systems in phenomenological thermodynamics is beyond the scope of this chapter. That states considered from the statistical viewpoint obey (CH) is shown in the next section directly, without relying on any additional underlying axioms.

For any other states of a system, i.e., states in $\Gamma$ that are not necessarily equilibrium states, the following axioms will be assumed~\cite{Lieb2013}.
\begin{enumerate}[(N1)]
\item Axioms (E1), (E2), (E3), and (E6), with $Z_0$ and $Z_1 \in \EqSt$ in axiom (E6), hold for all states in $\Gamma$.
\item  For any $X \in \Gamma$ there exist $X_0$ and $X_1 \in \EqSt$ with $X_0 \prec X \prec X_1$.
\end{enumerate}

Axiom (N2) specifies that for any non-equilibrium state of a system to be considered there should exist some equilibrium state from which it can be generated by means of an adiabatic process and that each non-equilibrium state can be brought to equilibrium with an adiabatic process. In the case of a box of gas, this could be achieved by letting the system equilibrate.

It is possible to assign values to the states of a system that quantify their use as a resource for performing tasks with  adiabatic operations. Intuitively, a state $X \in \Gamma$ is more valuable than another state $Y \in \Gamma$ if it allows for the generation of $Y$ with an adiabatic operation. More precisely, if the two states can be compared with $\pmac$, then $X$ is more valuable than $Y$ if and only if $X \pmac Y$ and $Y \not\pmac X$. The assignment of values to states should thus reflect this order.\footnote{In general, $\pmac$ may not be a total preorder and the ordering of values assigned to states that cannot be compared by means of $\pmac$ may be ambiguous.}

Lieb and Yngavson defined entropy functions as such a value assignment. They show that there is a (essentially) unique real valued function $S$ on the space of all equilibrium states of a thermodynamic system that is additive, i.e., for any two states $X \in \EqSt$ and $X' \in \EqSt'$, $S((X,X'))=S(X)+S(X')$, extensive, i.e., for any $\alpha > 0$ and any $X \in \EqSt$, $S(\alpha X)=\alpha S(X)$, and monotonic, i.e., for two states $X$, $Y \in \EqSt$ that are related by means of $\prec$, $X \prec Y$ holds if and only if $ S(X) \leq S(Y)$. Hence, the lower the entropy, the more valuable the state in this resource theory.

\begin{thm}[Lieb \& Yngvason]
  Provided that Axioms (E1) to (E6) as well as (CH) are obeyed on ${\alpha \EqSt} \times {(1-\alpha) \EqSt}$ for any $0 \leq \alpha \leq 1$, 
  there exists a function $S$ that is additive under composition, extensive in the scaling and monotonic with respect to $\pmac$. 
  Furthermore, this function $S$ is unique up to a change of scale $C_1 \cdot S + C_0$ with $C_1 > 0$. 
\end{thm}
For a state $X \in \EqSt$, the unique function $S$ is given as
\begin{equation} 
S(X)= \sup \left\{ \alpha \mid ((1-\alpha) X_0, \alpha X_1) \pmac X \right\}, \label{eq:entropy}
\end{equation}
where the states $X_0, X_1 \in \EqSt$ may be chosen freely as long as $X_0 \pmac X_1$ and $X_1 \not\pmac X_0$. This choice only changes the constants $C_0$ and $C_1$. In the case $X_\mathrm 0 \prec X \prec X_\mathrm 1$, \eqref{eq:entropy} can be intuitively understood as the optimal ratio $(1-\alpha) : \alpha$ of $X_0$ and $X_1$ such that the state $X$ can be created in an adiabatic process.\footnote{This definition extends to states obeying $X \prec X_0$ or $X_1 \prec X$~\cite{Lieb1999}, where it has to be interpreted slightly differently. To see this, note that the expression
\begin{equation} 
((1-\alpha) X_0, \alpha X_1) \prec X
  \label{eq:expression-in-entropy-X0X1precX}
\end{equation}
 is equivalent to 
$(((1-\alpha) X_0, \alpha X_1), \alpha' X_1) \prec (X, \alpha'
X_1)$ 
for any $\alpha'\geq 0$ and hence also to
$((1-\alpha) X_0, (\alpha'+\alpha) X_1) \prec (X, \alpha' X_1)$.  This
allows us to consider negative $\alpha$ (while only using positive scaling factors). For $\alpha<0$, the
condition~\eqref{eq:expression-in-entropy-X0X1precX} is to be understood as a transformation
$((1-\alpha) X_0) \prec (X, |\alpha| X_1)$ (choosing $\alpha'=-\alpha$).  A similar argument shows that for $\alpha>1$ the condition $((1-\alpha) X_0, \alpha X_1) \prec X$ can be understood as $\alpha X_1 \prec ((\alpha-1) X_0, X)$.}

Lieb and Yngvason derive equilibrium thermodynamics from the above axioms for equilibrium states and a few additional physically motivated properties, which also specify the geometric structure of the state space of thermodynamic systems a little more (for instance introducing convex combinations of states)~\cite{Lieb1998,Lieb1999,Lieb2001}. Their considerations rely on the phenomenological description of systems (by referring to simple systems for instance). We refer to~\cite{Thess} for a textbook on thermodynamics formulated in this approach. 
In addition, Lieb and Yngvason derive bounds on any monotonic extensions of $S$ from $\EqSt$ to $\Gamma$ under the condition that the above axioms for non-equilibrium states hold~\cite{Lieb2013}. We consider here the following slight adaptations of their bounds\footnote{The bounds given in~\cite{Lieb2013} are 
\begin{alignat}{4} 
S_{\mathrm{-}}(X) &= \sup &&\left\{ S(X') \right. &&\mid \left. X' \in \EqSt , \  X' \pmac X \ \right\}& \  \label{eq:ly1} \\
S_{\mathrm{+}}(X) &= \inf &&\left\{ S(X'') \right. &&\mid \left. X'' \in \EqSt , \  X \pmac X'' \right\} & \ .  \label{eq:ly2}
\end{alignat}
In phenomenological thermodynamics, where equilibrium states are traditionally described in terms of continuous parameters (such as the internal energy and the volume for instance), there exists an equilibrium state  $X_\alpha$ for each $((1-\alpha) X_0, \alpha X_1)$ that obeys $X_\alpha \sim ((1-\alpha) X_0, \alpha X_1)$. Under these circumstances the bounds \eqref{eq:lymin} and \eqref{eq:lymax} coincide with \eqref{eq:ly1} and \eqref{eq:ly2} respectively. This is, however, not implied by the axioms and may not hold if systems are described in a different way (e.g.\ in the statistical approach taken in Section~\ref{sec:noisy}). Note also that for \eqref{eq:ly1} and \eqref{eq:ly2} the inequality in \eqref{eq:c1} need not be strict.}
\begin{alignat}{2} 
S_{\mathrm{-}}(X) &= \sup &&\left\{ \alpha \mid ((1-\alpha) X_0, \alpha X_1) \pmac X \right\}  \label{eq:lymin} \\
S_{\mathrm{+}}(X) &= \inf &&\left\{ \alpha \mid X \pmac ((1-\alpha) X_0, \alpha X_1) \right\}. \label{eq:lymax} \end{alignat}
$S_-$ specifies the maximal fraction of the system that may be in state $X_1$ if $X$ is formed by combining systems in states $X_0$ and $X_1$ scaled appropriately, giving a measure for the resources needed to produce $X$. $S_+$ is the minimal portion of $X_1$ that is recovered from $X$ when transforming it into a system composed of scaled copies of $X_0$ and $X_1$. The difference $S_-(X) - S_+(X) \leq 0$ can be viewed as a measure for the resources that are used up when producing $X$ and decomposing it again (quantified in proportion of $X_0$ and $X_1$). For equilibrium states, where $S_-(X)=S_+(X)=S(X)$, this can be achieved without generating any overall resource loss.

$S_-$ and $S_+$ also provide necessary conditions as well as sufficient conditions for state transformations between non-equilibrium states by means of adiabatic processes.

\begin{prop}[Lieb \& Yngvason]\label{prop:necsuff}
Let $X,~Y \in \Gamma$. Then, the following two conditions hold:
\begin{align} 
S_{+}(X) < S_{-}(Y) &\implies X \pmac Y \ , \label{eq:c1}\\
X \pmac Y &\implies S_{-}(X) \leq S_{-}(Y) \text{ and } S_{+}(X) \leq S_{+}(Y) \ .
\end{align}
\end{prop}
Note that these conditions are closely related to the conditions we shall derive in the error-tolerant setting in Proposition~\ref{prop:necsuff_error}.

\section{Adiabatic processes in the statistical approach} \label{sec:noisy}
In this section, we show that the axiomatic framework that was introduced to describe systems phenomenologically applies also to a statistical description of systems: The phenomenologically-motivated axioms are satisfied by an order relation that is based on a statistical description of systems and adiabatic processes affecting them. In this sense, we can see the axiomatic framework as overarching these two approaches to thermodynamics.   

Here, the state of a system is described in terms of a density function over the space of microscopic states. Classically this is a distribution over phase space. In the quantum case, which we consider here for generality, it is a density operator on a Hilbert space $\cH$, denoted $\rho \in \Gamma=\cS(\cH)$.
An adiabatic process, which leaves no trace on the environment except for a change in the relative position of a weight, affects the microscopic degrees of freedom in the following way~\cite{Weilenmann2015}. 

\begin{defn} \label{def:adiabatic}
An \emph{adiabatic process} maps a state $\rho \in \cS(\cH_S)$ to another state
\begin{equation}
\sigma=\partr{U (\rho \otimes \tau) U^{\dagger}}{A} \  , 
\end{equation}
where $\tau \in \cS(\cH_A)$ is a \emph{sharp state}, meaning a state for which all its non-zero eigenvalues are equal, $U$ is an arbitrary unitary and the partial trace is taken over the subsystem $A$, where it is further required that 
\begin{equation} \label{eq:weight}
\partr{U (\rho \otimes \tau) U^{\dagger}}{S}=\tau \ . 
\end{equation}
\end{defn}

The evolution with an arbitrary unitary $U$ (that does not have to commute with the Hamiltonian) is enabled by the weight system, which is assumed to contain coherence as a resource that can be used catalytically. We rely on a model introduced in~\cite{Aberg2014}, where the weight is in a  superposition of energy eigenstates of a Hamiltonian with equally spaced energy levels. These may extend to infinity or (more physically) be bounded from below. Such a model could for instance be practically  realised with a laser, which naturally emits light in a coherent state. There are other weight models that can be used and that lead to similar behaviour; for instance, a weight modelled through switching potentials on and off as considered in~\cite{Allahverdyan2004} also leads to the emergence of arbitrary  unitary operations on the system (and environment $\tau$). Intuitively, the weight can be understood as a work storage system that is able to supply or absorb any amount of work, eliminating the significance of energetic considerations for these processes from the framework. (For further details regarding the weight system, we refer to~\cite{Aberg2014,Weilenmann2015}.) 

The state $\tau$ represents the part of the environment that is affected by the interaction with the weight system. The condition \eqref{eq:weight} encodes the requirement that the operations leave no trace on the environmental system after the interaction (except for the change in the weight). 
The sharp states take the role of the equilibrium states of the framework.
This is inspired by the fact that the microcanonical states have this property. However, here we apply our approach to systems of any sizes, including microscopic systems, and a notion of equilibration in terms of time evolution is not necessarily available. Nonetheless, there is a class of states that are equilibrium states in the sense that they obey all of the corresponding axioms, namely (E1)--(E6) and (CH). These are precisely the sharp states.

The preorder relation induced by the adiabatic processes of Definition~\ref{def:adiabatic} on the state space $\Gamma= \mathcal{S}(\mathcal{H})$ is closely related to the mathematical notion of majorisation (as stated in Proposition~\ref{prop:majo} below).

\begin{defn}
Let $\rho$, $\sigma \in \cS(\cH)$ with Hilbert space dimension $\dim{(\cH)}=n$ be two states with spectra $\left\{\ithEV{\rho} \right\}_i$ and $\left\{\ithEV{\sigma} \right\}_i$ respectively, ordered such that $\ithEVk{\rho}{1} \geq \ithEVk{\rho}{2} \geq \ldots \geq \ithEVk{\rho}{n}$ and $\ithEVk{\sigma}{1} \geq \ithEVk{\sigma}{2} \geq \ldots \geq \ithEVk{\sigma}{n}$. Then $\rho$ \emph{majorises} $\sigma$, denoted $\rho \maj \sigma$,\footnote{We alert the reader to the non-standard notation for the order relation (see also Footnote~\ref{fnt:notation}).} if for all $1 \leq k \leq n$,
\begin{equation}
\sum_{i=1}^{k} \ithEV{\rho} \geq \sum_{i=1}^{k} \ithEV{\sigma} \ .
\end{equation}
\end{defn}
Mathematically, $\maj$ is a \emph{preorder}, meaning it is reflexive, and transitive. The following proposition specifies its relation to adiabatic processes, which was proven in~\cite{Weilenmann2015} based on results concerning the resource theory of noisy operations~\cite{Horodecki2003,Horodecki2003b,Gour2013}.

\begin{prop} \label{prop:majo}
For $\rho$, $\sigma \in \cS(\cH)$, $\rho \maj \sigma$ if and only if $\rho$ can be transformed into $\sigma$ by an adiabatic operation.
\end{prop}

\begin{ex}\label{ex:adiabstat}
Consider the Helium gas from Example~\ref{ex:adiab}, where $\rho_X=\frac{\Pi_{\Omega_\mathrm{micro}(U,V)}}{\Omega_\mathrm{micro}(U,V)}$ and $\rho_Y=\frac{\Pi_{\Omega_\mathrm{micro}(2U,V)}}{\Omega_\mathrm{micro}(2U,V)}$ with $\Omega_\mathrm{micro}(U,V)$ the microcanonical partition function and $\Pi_{\Omega_\mathrm{micro}(U,V)}$ the projector onto its subspace. Then the two states obey $\rho_X \maj \rho_Y$.
\end{ex}

The composition of two systems characterised by density operators $\rho,~\sigma \in \Gamma$ is defined as their tensor product (replacing the Cartesian product considered in the phenomenological setting), i.e., $(\rho,~\sigma)$ means $\rho \otimes \sigma$ here. 
Scaling a system in a sharp state (corresponding to an equilibrium state in the axiomatic framework) by a factor $\alpha \in \mathbbm{Z}_{\geq 0}$ corresponds to taking its tensor power, $\alpha \rho=\rho^{\otimes \alpha}$. This operation can be formally extended to arbitrary $\alpha \in \mathbbm{R}_{\geq 0}$, which physically involves the consideration of processes on a larger system. For the details of this we refer to~\cite{Weilenmann2015}.

The following proposition relates the statistical description of adiabatic processes to the phenomenological approach introduced in Section~\ref{sec:resource}.

\begin{prop} \label{prop:axioms_LYmicro}
Consider adiabatic operations on states in $\Gamma=\mathcal{S}(\mathcal{H})$, with an equilibrium state space, $\EqSt$, made up of all sharp states. Then, the axioms (E1) to (E6), (CH) and (N1) and (N2) are obeyed. The corresponding entropy functions (according to \eqref{eq:lymin} and \eqref{eq:lymax}) are
\begin{align} 
S_-(\rho)&= H_\mathrm{min}(\rho) \label{eq:nonsmoothS} \\
S_+(\rho)&= H_0(\rho) , \label{eq:nonsmoothS2}
\end{align}
where $H_\mathrm{min}(\rho)=-\log \maxEV(\rho)$ and $\maxEV(\rho)$ is the maximal eigenvalue of $\rho$, and where $H_0(\rho)= \log \rank ( \rho )$.
\end{prop} 
These entropies are generally known as \emph{min} and \emph{max entropies} and have initially been introduced in information theory. They have various applications, e.g., for characterising extractable randomness and the compressibility of data, respectively~\cite{Renner2005, Koenig2009IEEE_OpMeaning}. For $\rho \in \EqSt$ these entropies coincide and they coincide with the unique additive and extensive entropy $S$. This establishes a connection between  
 entropy in phenomenological thermodynamics and in the statistical approach by means of a general and rigorous framework. Note that the connection established by Proposition~\ref{prop:axioms_LYmicro} is different from Jaynes' work~\cite{Jaynes1957, Jaynes1957_2}, which connects entropy in information theory to entropy in statistical physics (rather than phenomenological thermodynamics).

For microscopic systems, the question of whether there exists an (idealised) adiabatic operation achieving a transition between two (exact) microscopic states may not be meaningful in realistic situations where, due to experimental limitations, certain states may not be experimentally distinguishable, or in situations where one is satisfied with obtaining the desired states approximately.\footnote{Notice that approximations are also relevant for macroscopic systems. However, we do not usually explicitly mention them there, as they are extremely accurate.}
To consider such situations, we now introduce an extended framework that can tackle such approximations. It furthermore does not rely on a continuous scaling operation, which (even though possible) is unnatural for microscopic systems. The reason is that processes that involve a scaling with non-integer factors often have to be physically interpreted as operations on a larger (potentially macroscopic) system~\cite{Weilenmann2015}.

\section{Axiomatic framework for error-tolerant resource theories}\label{sec:error}

In this section, we extend the axiomatic framework described above to take  approximations into account. This is important to make a resource theory practically useful. Indeed, it is often the case that an approximation of a desired output is good enough for an intended purpose, and, at the same time, much less costly (in terms of the required resources). For this reason, accounting for approximations is also usual in information theory. For example, in randomness extraction, allowing for a deviation from the desired uniformly random bit string usually enables the extraction of a much larger number of (close to) random bits from the same data.

Processes with error-tolerance could not be modelled in previously existing axiomatic frameworks. Error-tolerant frameworks differ in various fundamental ways from those earlier ones. An example illustrating this is the fact that, if two transformations, each with error-tolerance $\eps$, are composed (either sequentially or in parallel), then there may be no process with error-tolerance $\eps$ that achieves the overall transformation: the errors of the separate processes may add up and there is generally no alternative process with a smaller error. Error-tolerant resource theories can therefore not be described with a transitive order relation. Instead, an error-tolerant resource theory introduces a family of order relations, $\left\{ \prec^\eps \right\}_\eps $, one for each error-tolerance $\eps$:
for $X$, $Y \in \Gamma$ the relation $X \prec^\eps Y$ expresses that a transformation from $X$ to $Y$ with error at most $\eps$ is possible, $X \sim^\eps Y$ means that there is a process transforming $X$ to $Y$ as well as one transforming $Y$ to $X$. We require $\prec^\eps$ to satisfy a few natural axioms. 

\begin{enumerate}[({A}1)]
\item Reflexivity: For any state $X \in \Gamma$ and any $\eps \geq 0$, 
$X \prec^{\eps} X$.
\item Ordering of error-tolerances: For any $X$, $Y \in \Gamma$ and any $\eps'\geq \eps \geq 0$, $X \prec^\eps Y \implies X \prec^{\eps'} Y$.
\item Additive transitivity: For any $X$, $Y$, $Z \in \Gamma$ and any $\eps$, $\delta \geq 0$, $X \prec^\eps Y$ and $Y \prec^\delta Z$ $\implies$ $X \prec^{\eps+\delta} Z$. 
\item Consistent composition: For any $X$, $Y \in \Gamma$, $Z \in \Gamma'$ and any $\eps \geq 0$, $X \prec^{\eps} Y$ $\implies$ $(X,Z)\prec^{\eps}(Y,Z)$.
\end{enumerate}

Axioms (A1), (A3) and (A4) are adaptations of (E1), (E2) and (E3) to the error-tolerant setting. In particular, they recover the latter axioms in the error free case ($\eps=0$). The Axiom (A2) further relates the different $\prec^\eps$, it expresses that increasing the error-tolerance increases the set of possible transformations. That such an error-tolerant description allows us to describe scenarios that could not be treated previously is illustrated with the following toy example.

\begin{ex}\label{ex:comp}
Consider a box containing a mole of Helium gas and let the states $X$ and $Z$ be given as in Example~\ref{ex:adiab}. Now assume that $X$ and $Z$ are used to encode information (they may for instance encode bit values $0$ and $1$ respectively) and that they are produced as the output of some computation that we know to yield either $X$ or $Z$ with probabilities $p$ and $(1-p)$ respectively. We describe the output state of our computation as $p X + (1-p) Z$, which corresponds to a state $(pU+(1-p)2U,pV+(1-p)2^{-\frac{3}{2}}V)$. This state can neither be transformed into $X$ nor $Z$ with an adiabatic operation. However, when considering error-tolerant adiabatic processes, an error probability of $\eps=\frac{1}{2}$ is sufficient to enable a transformation from $p X +(1-p) Z$ to $X$ (or $Z$) for any $p$. If $p \geq \frac{1}{2}$ this is achievable with the identity operation, if $p < \frac{1}{2}$, the adiabatic operation that transforms $Z$ to $X$ (and $X$ to another state) achieves this with error probability $\eps \leq \frac{1}{2}$. 
\end{ex}

The statistical viewpoint on this will be provided in Example~\ref{ex:comp2} below. There, we shall also quantify errors in a more rigorous manner, which will give us further insights.

\subsection{Quantifying resources in an error-tolerant framework} \label{sec:smoothentropies}

We are interested to quantify the value of the different states of a system as resources in an error-tolerant resource theory. In the error-free case, the relevant quantity is the position of the state in the state space with respect to the (transitive) preorder relation $\prec$. Tolerating an error of $\eps >0$ affects this value, making the same resource state more potent. 
This can be specified with the help of a \emph{meter system}, which is characterised by a single parameter (as it essentially serves to specify the position of states in the ordering $\prec$). We shall consider interactions between the system of interest and such a meter that simultaneously change the state of the system and of the meter, where the relative state change of the meter  provides information about the system's state.

In Lieb and Yngvason's framework the subspace of equilibrium states of a thermodynamic system is such a meter system~\cite{Lieb2014}. They have used this meter to specify the entropy of its own non-equilibrium states as well as that of the states of other systems whose state space does not have its own subspace of equilibrium states obeying the axioms, e.g.\ gravitating bodies~\cite{Lieb2014}. The meter systems we shall introduce in the following are more general in the sense that we do not impose a continuous scaling operation on their state space. Instead, we allow a meter to be characterised by a parameter that may be discrete (rather than continuous) and have a finite (rather than infinite) range.

We consider a meter system with state space $\Gamma_{\lambda}=\left\{ \chi_\lambda \right\}_{\lambda \in \Lambda}$, specified with a function
\begin{align*}
\chi:\ \Lambda&\to \Gamma_\lambda\\
\lambda&\mapsto\chi_\lambda
\end{align*}
with $\Lambda\subseteq\mathbb{R}_{\geq 0}$, where the parameter $\lambda$ labels the different states. The change in $\lambda$ produced during an interaction with a system shall specify the resource value of the system of interest.

From now on, we shall quantify errors in terms of probabilities, i.e., we shall require $0 \leq \eps, \delta \leq 1$. Whenever we add two errors we understand this as  $\eps+\delta \defeq \min \left\{ \eps+\delta, \  1 \right\}$. While the axiomatic framework does not rely on this restriction, the particular choice of the function $f(\eps)$ below that specifies the resource-error tradeoff is motivated by the probabilistic interpretation of errors. 

We assume a meter system to obey the following axioms. 
\begin{enumerate}[({M}1)]
\item Reduction property: For any $X$, $Y \in \Gamma$, for any meter state $\chi_\lambda \in \Gamma_\lambda$ and any $\eps \geq 0$,
$(X,\chi_\lambda) \prec^\eps (Y,\chi_{\lambda})$ $\implies$ 
$X \prec^\eps Y$.
\item Additivity of meter states: For any $\chi_{\lambda_1}$, $\chi_{\lambda_2} \in \Gamma_\lambda$, $\chi_\lambda \sim (\chi_{\lambda_1}, \chi_{\lambda_2}) \iff  \lambda=\lambda_1+ \lambda_2$.
\item Ordering of meter states: A meter system has at least two inequivalent states and its states are labelled monotonically in $\lambda$, such that
for any $\chi_{\lambda_1}$, $\chi_{\lambda_2} \in \Gamma_\lambda$,  
$\lambda_1 \leq \lambda_2 \iff \chi_{\lambda_1} \prec \chi_{\lambda_2}$.
\item Resource-error tradeoff: For any $\lambda_1 > \lambda_2 \in \Lambda$ and any $0 \leq \eps \leq 1$,
$\chi_{\lambda_1} \prec^\eps \chi_{\lambda_2} \implies  \lambda_1 \leq \lambda_2 + f(\eps)$ with $f(\eps)=- \log_2(1-\eps)$. 
\end{enumerate}

A meter system is supposed to measure the value of different resources. Axiom (M1) demands that it acts passively in the sense that it should not enable otherwise impossible state transformations. This property is also obeyed by the equilibrium states in Lieb and Yngvason's axiomatic framework, where it is  known as the cancellation law~\cite{Lieb1998,Lieb1999}. Axioms (M2) and (M3) concern merely the labelling of meter states; according to (M2) this should be additive under composition and according to (M3) monotonic with  respect to the error-free ordering $\prec$. Note that (M3) also ensures that all states $\chi_\lambda \in \Gamma_\lambda$ can be compared with $\prec$, i.e., for any such states $\chi_{\lambda_1}$, $\chi_{\lambda_2} \in \Gamma_\lambda$ either $\chi_{\lambda_1} \prec \chi_{\lambda_2}$ or $\chi_{\lambda_2} \prec \chi_{\lambda_1}$.

Axiom (M4) specifies the transformations on $\Gamma_\lambda$ that are enabled by accepting imprecisions.   
According to Axiom (A2), a higher error-tolerance cannot prohibit any transformations but may enable more. A bound on these is specified on the meter in terms of the function $f(\eps)$, which is non-decreasing in $\eps$ and obeys $f(0)=0$ (according to Axiom (M3)).   
Since we understand $\eps$ as an error probability such that $0 \leq \eps \leq 1$, any state transformation should be possible in the extreme case of $\eps=1$. Thus, in addition to $f(0)=0$, we require $\lim_{\eps \rightarrow 1}f(\eps)= \infty$ (as the allowed values for $\lambda$ can be unbounded if we allow for the composition of an arbitrary number of meter systems). Furthermore, it should always be possible to run $n$ independent instances of a process in parallel, in which case the success probabilities $(1-\eps)$ should multiply. In the case of a meter system we take it that there is also no alternative process with a lower error probability, i.e., we require that the process is possible if and only if the $n$ parallel instances of that process are possible. We thus require that the existence of a process $\chi_{\lambda_1}\prec^\eps \chi_{\lambda_2}$ implies that for any $n \in \mathbbm{N}_+$,
\begin{align}
n \cdot \lambda_1 &\leq n \cdot \lambda_2 +f(1-(1-\eps)^n) \ .
\end{align}
This can be ensured with $f(1-(1-\eps)^n)= n \cdot f(1-(1-\eps))$, which implies that $f(\eps)= - c \cdot \log_2(1-\eps)$ (assuming continuity of $f$). We choose $c=1$ for simplicity, which fixes a scale for the parameter $\lambda$.
 According to (M4), increasing the error-tolerance allows for an increase in $\lambda_1 -\lambda_2$,  we therefore call this axiom a resource-error tradeoff.

\begin{defn} \label{def:meter}
A meter system with state space $\Gamma_\lambda$ is \emph{suitable}
for measuring a system with state space $\Gamma$ if it obeys (A1) to (A4) and the axioms for meter systems (Axioms (M1) to (M4)), and, if there exists a reference state denoted by $\refStph \in \Gamma$ such that for any  state $X \in \Gamma$  there exist meter states  $\chi_{\lambda_1(X)}$, $\chi_{\lambda_2(X)}$, $\chi_{\lambda_3(X)}$, $\chi_{\lambda_4(X)} \in \Gamma_\lambda$ such that $(\refStph, \chi_{\lambda_1(X)}) \prec (X, \chi_{\lambda_2(X)})$ and $(X, \chi_{\lambda_3(X)}) \prec (\refStph, \chi_{\lambda_4(X)})$ hold. 
\end{defn}

Relying on the notion of a suitable meter system, we define the following quantities. 

\begin{defn} \label{definition:smoothentropies}
For an error-tolerant resource theory with state space $\Gamma$ and a suitable meter system $\Gamma_\lambda$ we define for each $\eps \geq 0$ and for each $X \in \Gamma$,
\begin{alignat}{2}
S^\epsilon_-(X)&\defeq \sup &&\left\{\lambda_1-\lambda_2 \ \middle| \ (\refStph,\chi_{\lambda_1})\prec^\epsilon (X,\chi_{\lambda_2})\right\} \  \\
S^\epsilon_+(X)&\defeq \inf &&\left\{\lambda_2-\lambda_1 \ \middle| \ (X,\chi_{\lambda_1})\prec^\epsilon (\refStph,\chi_{\lambda_2})\right\} \ ,
\end{alignat}
where $\refStph \in \Gamma$ is a fixed reference state and $\chi_{\lambda_1}$, $\chi_{\lambda_2} \in \Gamma_\lambda$. 
\end{defn}

Due to the suitability of the meter system, $S^\eps_-$ and $S^\eps_+$ are defined for all $X \in \Gamma$. 
In terms of these quantities, we can derive necessary conditions and sufficient conditions for state transformations. In the error-free case, i.e., if $\eps=\eps'=\delta=0$ in Proposition~\ref{prop:necsuff_error} below, the relations from Proposition~\ref{prop:necsuff} are recovered for $S_-^0$ and $S_+^0$.

\begin{prop} \label{prop:necsuff_error}
Consider an error-tolerant resource theory with state space $\Gamma$ that obeys Axioms~(A1) to (A4) and a suitable meter system. Then for any $X$, $Y \in \Gamma$ and $\eps, \eps' \geq 0$, the following conditions hold:  
\begin{align}
S_+^\eps(X) < S_-^{\eps'}(Y) &\implies X \prec^{\eps+\eps'} Y \ , \\
X \prec^{\eps} Y &\implies   S_-^{\delta}(X) \leq S_-^{\delta+\eps}(Y)  \text{ and }  S_+^{\delta+\eps}(X) \leq S_+^{\delta}(Y) \text{ for any } \delta \geq 0 \ .
\end{align}
\end{prop}

Because $\prec^{\eps}$ is intransitive for $\eps>0$, $S^\eps_-$ and $S^\eps_+$ are not monotonic with respect to $\prec^{\eps}$ (except for $\eps=0$). For the proof of this proposition and further properties of $S_-^\eps$ and $S_+^\eps$ , for instance the respective super- and sub-additivity expected for entropy measures, we refer the reader to Propositions~7.3.8 and 7.3.9 as well as the subsequent elaborations in~\cite{thesis}.

\section{Error-tolerant resource theories in the statistical approach}
\label{sec:microerror}
In this section we treat states in the usual quantum-mechanical framework. This means that the state space is $\Gamma= \cS(\cH)$, the set of density operators on a Hilbert space $\cH$. Composition of states is defined as their tensor product and adiabatic operations on these states were introduced in Definition~\ref{def:adiabatic}. 

We will rely on a meter system for which the meter states $\chi_\lambda \in \Gamma_\Lambda$ have eigenvalues $2^{-\lambda}$ and $0$ with multiplicities $2^\lambda$ and  $\dim (S)- 2^\lambda$ respectively, where $\dim (S)$ is the dimension of the Hilbert space $\chi_\lambda$ acts on, in accordance with~\cite{Gour2013,Weilenmann2015,Kraemer2016}. The spectrum of such a state can be conveniently written as a step function, 
\begin{equation} \label{eq:meter}
 f_{ \chi_\lambda}(x) = 
\begin{cases} 
2^{-\lambda} & x \leq 2^\lambda \ ,\\
0 & 2^\lambda < x \leq \dim (S) \ .
\end{cases}
\end{equation}
We let the parameter $\lambda$ take values $\lambda=\log_2(n)$ for $n=1, \ldots, \dim{(S)}$, where $\dim{(S)}>1$  and we assume that a meter system $\Gamma_\Lambda$ with arbitrarily large dimension, $\dim (S)$, may be chosen. In the definition of $S_-^{\eps}$ and $S_+^{\eps}$ we shall optimise over  meter systems of different sizes.  
We will compare this to another meter system in Example~\ref{ex:battery}, a model which is also known as a \emph{battery}~\cite{Bennett1982}.

\label{sec:smoothingNO}

In a thermodynamic process, imprecisions may occur and be tolerated either in the input states to a process, its outputs, or both. Imprecise input states account for errors in the preparation of a system. When tolerating such errors, an error-tolerant transition is possible if a transition from an approximate input state to the target output can be achieved. In case of imprecisions in the output, an error-tolerant transformation is possible if a transformation could reach the output state approximately.
These two types of errors can moreover be combined to an overall error $\eps$. We shall call these three ways of quantifying errors \emph{smoothings}. They were previously considered in the context of resource theories in~\cite{Horodecki2017,VanderMeer2017,Hanson2017,Hanson2017a,thesis}. When quantified by means of the generalised trace distance, these smoothings are equivalent for adiabatic operations, meaning that they all lead to the same family of order relations $\left\{ \maj^\eps \right\}_\eps$ (adaptations of the majorisation relation). The equivalence of these three smoothings was proven in~\cite{thesis, Horodecki2017}, allowing us (without loss of generality) to state only one of the equivalent definitions in the following.

\begin{defn} \label{def:trdist}
The resource theory of \emph{smooth adiabatic operations} is characterised by the order relations $\left\{ \maj^\eps \right\}_\eps$, where
$\rho\maj^\epsilon \sigma$ if and only if $\exists \ \rho',\sigma' \text{ and } \epsilon', \epsilon'' \text{ s.t. } \rho'\maj \sigma'\text{ and } \rho'\in \mathcal{B}^{\epsilon'}(\rho),\  \sigma'\in\mathcal{B}^{\epsilon''}(\sigma)$ with $\epsilon'+\epsilon''\leq\epsilon$; $\epsball{\eps}{\rho} \defeq \left\{ \rho' \in \cS(\cH)  \mid \trdist{\rho}{\rho'} \leq \eps \right\}$ denotes the set of all states that are $\eps$-close to the state $\rho \in \cS(\cH)$, measured in terms of the trace distance $\trdist{\cdot}{\cdot}$~\cite{TomamichelBook}.\footnote{For $\eps=0$ the usual majorisation relation is recovered.}
\end{defn}

The following example illustrates that resource theory of smooth adiabatic operations allows us to analyse situations that could not be described within the resource theory of adiabatic processes. It is analogous to Example~\ref{ex:comp}, but described from a statistical perspective. 
\begin{ex} \label{ex:comp2}
Consider once more a mole of Helium gas with states $\rho_X=\frac{\Pi_{\Omega_\mathrm{micro}(U,V)}}{\Omega_\mathrm{micro}(U,V)}$ and $\rho_Z=\frac{\Pi_{\Omega_\mathrm{micro}(2U,2^{-3/2}V)}}{\Omega_\mathrm{micro}(2U,2^{-3/2}V)}$. Now assume again that these states are used to encode information (they may for instance encode bit values $0$ and $1$ respectively) and that they are produced as the output of a computation that we know to yield either $\rho_X$ or $\rho_Z$ with probabilities $p$ and $(1-p)$ respectively. We describe the output state of this computation as $p \rho_X + (1-p) \rho_Z$, which cannot be transformed into $\rho_X$ (or $\rho_Z$) with an adiabatic operation. However, when considering error-tolerant adiabatic processes, an error-probability of $\eps=\frac{1}{2}$ is sufficient to enable a transformation from $p \rho_X +(1-p) \rho_Z$ to $\rho_X$ for any $p$ (as $p \rho_X +(1-p) \rho_Z \maj^\eps \rho_X$). 
More generally, the required error probability to achieve this transformation   is $\eps= \min \left\{p, 1-p \right\}$. For smaller $\eps$ additional resources would be required to recover $\rho_X$, which can be seen as an example of the resource-error tradeoff.
\end{ex}

In the following we derive the quantities $S_-^\eps$ and $S_+^\eps$ that provide necessary conditions as well as sufficient conditions for the existence of smooth adiabatic operations between different states according to Proposition~\ref{prop:necsuff_error}.

\begin{prop}
\label{prop:NO_axioms}
The resource theory of smooth adiabatic operations with state space $\Gamma=\cS(\cH)$ and with composition of states defined as their tensor product obeys Axioms (A1) to (A4). The meter system $\Gamma_\Lambda$ defined by~\eqref{eq:meter} is suitable for measuring systems with a state space $\Gamma=\cS(\cH)$. With reference state $\refSt=\ketbra{0}{0} \in \cS(\cH)$, we obtain
\begin{align}
S^\epsilon_-(\rho)&=\Hmineps{\eps}{\rho} \  \label{eq:sma1} \\
S^\epsilon_+(\rho)&=\HHeps{1-\epsilon}{\rho} + \log_2(1-\eps) \ , \label{eq:sma2}
\end{align}
where
\begin{align}
\Hmineps{\eps}{\rho} &\defeq \sup_{\rho' \in \epsballsub{\eps}{\rho}} \Hmin{\rho'} \  \\
\HHeps{1-\eps}{\rho} &\defeq \log \inf \left\{ \frac{1}{1-\eps} \tr{Q} : \tr{Q \rho}\geq 1-\eps \ \mathrm{ and } \ 0\leq Q \leq \mathbbm{1} \right\} \ .
\end{align} 
\end{prop}
This is shown by means of Proposition~7.2.7, Lemma~7.3.12 and Proposition~7.3.13 in~\cite{thesis}. $\HminepsO{\eps}$ is known as a \emph{smooth min entropy} and $\HHepsO{1-\eps}$ as a \emph{smooth max entropy} in single-shot information theory~\cite{Dupuis2013_DH}. 
Note further that for $\eps=0$ we recover the quantities $S_-$ and $S_+$ from Proposition~\ref{prop:axioms_LYmicro}.

Alternatively, we may consider other meter systems that are less fine-grained. An example is a meter system that consists of a number of small systems in two possible states, one of which is a resource state. The number of such resource states that is consumed (or gained) in the construction or destruction of a particular state is then a measure for the resourcefulness of that state. This type of meter system is also known as a \emph{battery} in the literature, going back to ideas of Bennett~\cite{Bennett1982}.

\begin{ex}\label{ex:battery}
Take the meter system to be an arbitrarily large collection of qubits that can each be either in a pure state $\rho=\ketbra{0}{0}$ or in a maximally mixed state $\rho=\frac{\id_2}{2}$. Thus, the spectrum of such a meter state $\chi_\lambda \in \Gamma_{\Lambda}$ can be written as a step function 
\begin{equation}
 f_{ \chi_\lambda}(x) = 
\begin{cases} 
2^{-\lambda} & x \leq 2^\lambda \ ,\\
0 & 2^\lambda < x \leq \dim (S) \ ,
\end{cases}
\end{equation}
where $\lambda \in \Z_+$ and $\dim (S)$ is the dimension of the Hilbert space $\chi_\lambda$ acts on. This meter system leads to coarse-grained entropy measures compared to the previously considered one, which can only take integer values. More precisely, it leads to
\begin{align}
S^\epsilon_-(\rho)&= \floor{ \Hmineps{\eps}{\rho} } \   \\
S^\epsilon_+(\rho)&= \ceil{ \HHeps{1-\epsilon}{\rho} + \log_2(1-\eps) } \ . 
\end{align}
With such a meter system, larger classes of states $\rho$ yield the same $S^\epsilon_-(\rho)$ (and similar for $S^\epsilon_+(\rho)$), i.e., the meter system is not fine-grained enough to distinguish them. 
For a qubit in a state $\rho$, for instance, $S_-^\eps(\rho)=0$ for all $\rho$ with $\frac{1}{2} < \maxEV\left(\rho\right)-\eps < 1$, where $\maxEV\left(\rho\right)$ is the larger eigenvalue of $\rho$.
\end{ex}

Our framework also exhibits the phenomenon of embezzling, which is  known from other resource theories~\cite{VanDam2003, Ng2014, Brandao2015b}. We define a catalytic transformation as one that can be achieved with any catalyst. For instance, there is a \emph{catalytic smooth adiabatic process} from a state $\rho$ to $\sigma$, denoted as $\rho \prec_\mathrm{cat}^\eps \sigma$, if and only if there exists a catalyst $\rho_\mathrm{cat}$ (a state on a finite dimensional Hilbert space) such that $\rho \otimes \rho_\mathrm{cat} \maj^\eps \sigma \otimes \rho_\mathrm{cat}$. In principle, by suitably engineering $\rho_\mathrm{cat}$ (allowing it to be an arbitrarily large system) we can achieve a transformation between any states $\rho$ and $\sigma$. To show this, it is sufficient to consider transformations on the meter systems. It turns out that, provided the tolerated failure probability $\eps$ is strictly positive, any transformation on the meter system is enabled by a suitable catalyst~\cite{thesis}.

\begin{prop} \label{prop:embezzling} 
For any meter state $\chi_\lambda \in \Gamma_\Lambda$ and for any $\eps > 0$ we can find a quantum state $\rho_\mathrm{cat}$ such that
\begin{equation}
\chi_{\lambda} \prec_\mathrm{cat}^\eps \chi_0 \ . 
\end{equation}
\end{prop}

\begin{figure}
\centering
\includegraphics[width=0.9\textwidth]{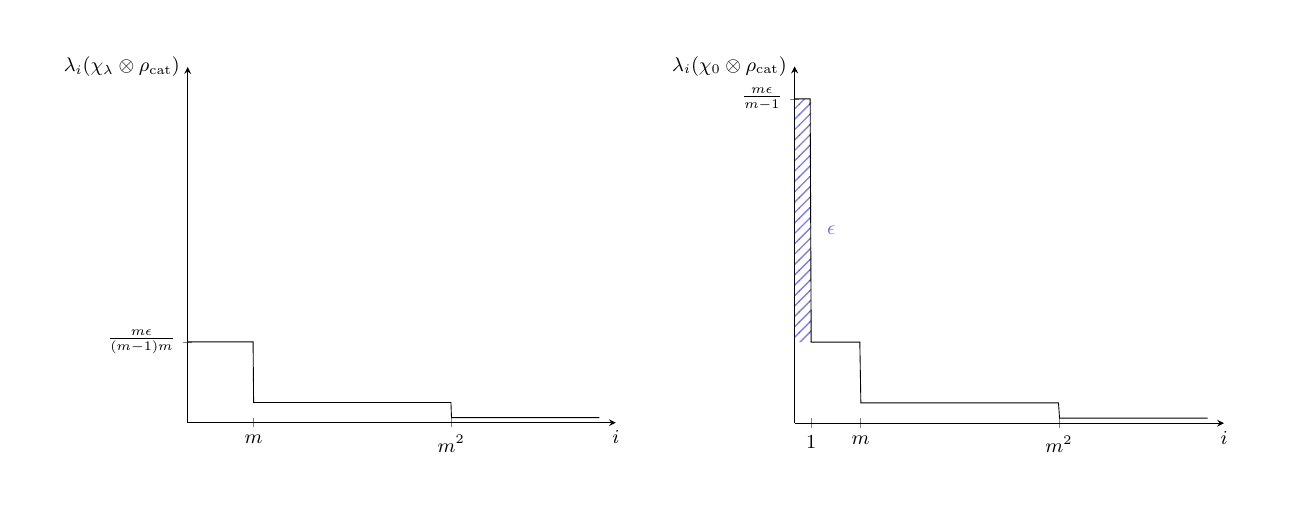}
\caption[.]{A transformation $\chi_\lambda \prec_\mathrm{cat}^\eps \chi_0$ is achieved with a catalyst $\rho_\mathrm{cat}$ with eigenvalues $\lambda_i(\rho_\mathrm{cat})= \frac{m \eps}{m-1} \cdot m^{-\ceil{\log_m i}}$, where $m=2^\lambda$ and where $i$ ranges from 1 to the to the maximal value that leaves $\rho_\mathrm{cat}$ normalised (and where the last eigenvalue may be smaller than prescribed so that normalisation is achieved). 
 On the left hand side we show the spectrum of $\chi_\lambda \otimes \rho_\mathrm{cat}$ and on the right right hand side that of $\chi_0 \otimes \rho_\mathrm{cat}$. They differ in the first eigenvalue (and in the last few, which is not visible in the plots), giving an overall trace distance of $\eps$.}
\label{fig:catalyst}
\end{figure}

Figure~\ref{fig:catalyst} illustrates how a corresponding catalyst can be engineered. Notice that since $\lambda > 0$ for all meter states (recall \eqref{eq:meter}) and according to (M3), $\chi_0$ is the most valuable meter state in the resource theory from which all others can be generated with an adiabatic process. 

That catalysts can enable any transformations may sound counter-intuitive at first: It is a phenomenon that, nonetheless, naturally occurs in experiments. Consider for instance an experiment in a laboratory, where we aim to excite an atom to a higher energetic state. The light present in the laboratory can be viewed as a large catalyst, that, if the atom is not properly isolated from it, may cause this transition without us perceiving any change in the laboratory's state. The part of this catalyst that is relevant for the transition is the light around a certain frequency, which we can model as a distribution over the number of photons at the corresponding energy, expressing our knowledge of the probability that $n$ photons of (approximately) that frequency are present. The absorption of one of these photons is enough to cause the transition, however, it changes our description of the catalyst only marginally (also without a supply of new photons).

If we were to impose restrictions on the type of system that can be used as a catalyst, for instance if in an experimental setup a system of a certain size is present, the processes the system could enable when considered as a catalyst would be more restrictive. This is illustrated with the following toy example.

\begin{ex}
Assume that we want to make the transformation $ \frac{\id_2}{2} \prec_\mathrm{cat}^\eps \ketbra{0}{0}$ and that we have only one additional qubit as a catalyst at our disposal. Let this qubit be $\rho$ with spectrum $(p, 1-p)$ where $p \geq \frac{1}{2}$. According to the majorisation condition,\footnote{We aim to achieve the transformation $\rho \otimes \frac{\id_2}{2} \maj^\eps \rho \otimes \ketbra{0}{0}$, where the ordered spectra of the two states are $(\frac{p}{2},\frac{p}{2},\frac{1-p}{2},\frac{1-p}{2})$ and $(p,1-p,0,0)$.} $\rho$ can be used as a catalyst that enables the transformation of $\frac{\id_2}{2}$ to $\ketbra{0}{0}$, if $\frac{1}{2}p \leq \eps$ and $1-p \leq \eps$. Hence, if $\eps < \frac{1}{3}$, a qubit catalyst is not sufficient.
\end{ex}

The size of the catalyst $\rho$, denoted $\dim(\rho)$, that is needed to enable a transformation
$\chi_{\lambda_1} \prec_{\mathrm{cat}}^\eps \chi_{\lambda_2}$ where $\lambda_2 < \lambda_1$ is bounded by means of the following necessary condition for such a transformation: for $\eps \dim(\rho) < 1$,\footnote{The largest eigenvalues of $\rho \otimes \chi_{\lambda_1}$ and $\rho \otimes \chi_{\lambda_2}$ are $p 2^{-\lambda_1}$ and $p 2^{-\lambda_ 2}$ respectively, where $p$ is the maximal eigenvalue of $\rho$. At rank $2^{\lambda_2}$, the sum of the eigenvalues of the two states are $p 2^{\lambda_2-\lambda_1}$ and $p$ respectively, thus the transition is only possible if $p 2^{\lambda_2-\lambda_1} \geq p-\eps$. }
\begin{equation}
\lambda_1-\lambda_2 \leq  -\log_2{(1-\eps \dim(\rho))} \ .
\end{equation}
For any given transformation on the meter system (i.e., for fixed $\lambda_1> \lambda_2$) this bound on the minimal catalyst size increases with decreasing error-tolerance $\eps$. The bound also shows that in the special case $\eps=0$ no transition with $\lambda_1 > \lambda_2$ is achievable.

Our axiomatic approach has the flexibility needed to describe different error types. We illustrate this in the following with the example of probabilistic transformations from~\cite{Alhambra2015}. A probabilistic transformation, $\cptp$, from a state $\rho \in \cS(\cH)$ to a state $\sigma \in \cS(\cH)$ with error probability $\eps$, is 
\begin{equation}
\rho \rightarrow \cptp(\rho)= (1- \eps) \sigma + \eps \xi,
\end{equation}
where $\xi$ is an arbitrary state. This expresses that the process, which aims to transform $\rho$ to $\sigma$, succeeds with probability $1-\eps$, whereas with probability $\eps$ any output can be produced. 

\begin{defn}\label{def:alvaro_smoothing}
The resource theory of \emph{probabilistic adiabatic operations} is characterised by the order relations $\left\{ \pmaj^\eps \right\}_\eps$, where for $\rho$, $\sigma \in \cS(\cH)$, 
$\rho \pmaj^{\eps} \sigma$ if and only if $\exists \xi \in \cS(\cH) \ \sth \rho \maj (1- \eps) \sigma + \eps \xi$.
\end{defn}

Probabilistic adiabatic transformations are characterised by $S_-^\eps$ and $S_+^\eps$ given in the following proposition, which provide necessary conditions and sufficient conditions for state transformations with such processes (according to Proposition~\ref{prop:necsuff_error}).

\begin{prop}
Probabilistic adiabatic transformations on a state space $\Gamma=\cS(\cH)$, where composition of states is defined as their tensor product, obey Axioms (A1) to (A4). The meter system $\Gamma_\Lambda$ defined by~\eqref{eq:meter} is suitable for measuring systems with a state space $\Gamma=\cS(\cH)$. With reference state $\refSt=\ketbra{0}{0} \in \cS(\cH)$, we obtain
\begin{align}
S_-^\eps(\rho)&=\Hmin{\rho}+ \log_2(1-\eps) \  \label{eq:sme1}\\
S_+^\eps(\rho)&=\Hzero{\rho} \ . \label{eq:sme2}
\end{align}
\end{prop}

This has been proven as Lemma~7.4.1 and Proposition~7.4.2 in~\cite{thesis}. We remark here that 
$\rho \pmaj^\eps \sigma$ and $\sigma \pmaj^\delta \omega$ imply $\rho \pmaj^{\eps+\delta-\eps \delta} \omega$, which for $\eps, \delta \neq 0$ is a strictly smaller error than the axiomatically required $\eps+\delta$.

The existence of a probabilistic adiabatic transformation from a state $\rho$ to a state $\sigma$ implies that there is also a smooth adiabatic transformation from $\rho$ to $\sigma$, hence the state transformations enabled by adiabatic probabilistic transformations are a subset of the smooth adiabatic ones. We can see this by considering the output state of a probabilistic adiabatic process (see Definition~\ref{def:alvaro_smoothing}), which obeys
\begin{equation}
\trdist{\sigma}{(1-\eps)\sigma+\eps \xi}= \eps \trdist{\sigma}{ \xi} 
\leq \eps \ ,
\end{equation}
because the trace distance of two states $\sigma$ and $\xi$ is bounded by $1$. The converse statement is not true. This can be seen by considering the states $\rho=\frac{1}{2}\ketbra{0}{0} + \frac{1}{2}\ketbra{1}{1}$ and $\sigma=\frac{3}{4} \ketbra{0}{0}+ \frac{1}{4} \ketbra{1}{1}$.  For an error-tolerance of $\eps=\frac{1}{4}$ there is a smooth adiabatic operation from $\rho$ to $\sigma$, since $\rho \maj^{\scaleto{1/4}{5.5pt}} \sigma$, but there is no corresponding probabilistic adiabatic process, $\rho \not\pmaj^{\scaleto{1/4}{5.5pt}} \sigma$. Instead, an error-tolerance of (at least) $\eps=\frac{1}{3}$ would be needed in the latter case.

\section{Axiomatic emergence of macroscopic thermodynamics}\label{sec:emergence}

In the limit of large systems our error-tolerant framework leads to the emergence of an effective order relation that is characterised by a single entropic quantity for thermodynamic equilibrium states, as we shall explain in the following. Even though our considerations in this and the subsequent section are based on insights from Chapter~8 of~\cite{thesis}, we take a slightly different approach here, where macroscopic states depend on a continuous parameter.

We first define the elements of a  macroscopic state space as functions of a continuous parameter $n \in \mathbbm{R}_{\geq 0}$ that map each $n$ to a state $X(n) \in \Gamma(n)$, i.e., to a state in a space $\Gamma(n)$ in a set of  state spaces $\left\{ \Gamma(n) \right\}$. The set of all such macroscopic states is
\begin{equation}
\Gamma_\infty \defeq \left\{ X_\infty  \  \middle| \begin{array}{lll} X_\infty:& \mathbbm{R}_{\geq 0} &\rightarrow \left\{\Gamma(n) \right\}  \\  &n  &\mapsto X(n) \in \Gamma(n) \end{array} \right\} \ .
\end{equation}
One may think of $n$ as encoding the amount of substance in a system (and the fact that it can take any value in a continuum corresponds to the usual approximation made in macroscopic thermodynamics). The states $X_\infty$ that intuitively describe a physical system are those for which $X(n')$ is the state of a subsystem of $X(n)$ for all $n' \leq n$. Among them, the equilibrium states are those that have essentially the same macroscopic  properties for all $n$, meaning that these properties scale linearly with the parameter $n$. Thinking for instance of the state of a gas in a box, a function that characterises such an equilibrium state is $X_\infty$ such that $X(n)= (nU,nV,nN)$ for all $n \in \mathbbm{R}_{\geq 0}$, where $nU$ is the internal energy, $nV$ the volume and $nN$ the matter content of a system. In order to introduce these intuitive notions into the formalism, we require a little more terminology and another axiom, which may be added to any error-tolerant resource theory.

\begin{enumerate}[({A}5)]
\item Let $\Gamma$ be the state space of the resource theory and $\Gamma_\lambda$ a meter system. Then there exists a state $Z \in \Gamma$ that for any $X$, $Y \in \Gamma$ obeys
\begin{equation} \label{eq:first}
X \prec^\eps Y  \ \iff \   (X, Z) \prec^\eps (Y, Z) 
\end{equation} 
and that for any meter states $\chi_{\lambda_1}$, $\chi_{\lambda_2} \in \Gamma_\lambda$ obeys\footnote{This condition is independent of~\eqref{eq:first} only if  $\Gamma_\lambda \not\subseteq \Gamma$, if $\Gamma_\lambda$ is chosen such that  $\Gamma_\lambda \subseteq \Gamma$ it is implied by~\eqref{eq:first}.}
\begin{equation}
\chi_{\lambda_1} \prec^\eps \chi_{\lambda_2}  \ \iff \   (\chi_{\lambda_1}, Z) \prec^\eps (\chi_{\lambda_2}, Z) \ . 
\end{equation} 
\end{enumerate}

The axiom encodes the requirement that a system should have at least one state  that cannot be used as a catalyst with respect to the resource theory.   
It is conceivable that a ground state of a system should generally have this property. Taking a statistical viewpoint, there are usually many such states. For example, in the quantum resource theories considered before, the maximally mixed state for adiabatic processes (and the meter states in $\Gamma_\Lambda$) and the Gibbs states for thermal operations have this property. 
From now on, whenever (A5) holds, we choose the reference state on the system, $\refStph \in \Gamma$ (recall Definition~\ref{def:meter}), such that it has this property. This choice fixes the zero of the entropic quantities we shall consider below. We then define the following subset of $\Gamma$, 
\begin{equation}\label{eq:setT}
\Gamma_\mathrm{M}= \left\{ \tau_\lambda \in \Gamma \ \middle| \ S_-^0(\tau_\lambda)= S_+^0(\tau_\lambda)=\lambda \right\} \subseteq \Gamma \ .
\end{equation}
From our choice of $\refStph$, it is easy to see that the set $\Gamma_\mathrm{M}$  contains at least this state, which obeys $S_-^0(\refStph)= S_+^0(\refStph)=0$, i.e., $\tau_0=\refStph$. The label $M$ stands for meter, as the states $\tau_\lambda \in \Gamma_\mathrm{M}$ behave like meter states, i.e., they obey Axioms (M1) to (M4).\footnote{Axiom (M1), for instance, follows since  $(X,\tau_\lambda) \prec^\eps (Y,\tau_\lambda)$ implies $(X,\tau_\lambda,\chi_{\lambda_1},\chi_{\tilde{\lambda_2}}) \prec^\eps (Y,\tau_\lambda,\chi_{\lambda_1},\chi_{\tilde{\lambda_2}})$ by (A4) and since by applying \eqref{eq:setT} and the property (A5) for $\refStph$ this can be shown to also  imply $X \prec^\eps Y$. (M2), (M3) and (M4) also follow from \eqref{eq:setT} and the axioms.}  
Macroscopic states made up of states $\tau_{\lambda_n} \in \Gamma_\mathrm{M}(n)$ are denoted as $\tau_{\lambda_\infty}\in \Gamma_\mathrm{M}^{\infty}$. 

On $\Gamma_\infty$, we introduce an effective macroscopic order relation $\prec_\infty$: we write $X_\infty \prec_\infty Y_\infty$ if for any $\eps >0$ there exists $n_0$ such that for all $n \geq n_0$, $X(n) \prec^{\eps} Y(n)$.    
Intuitively, equilibrium states have the property that their behaviour with respect to $\prec_\infty$ is essentially characterised by a single quantity which we call $\lambda_{X_\infty}$, i.e., an equilibrium state $X_\infty$ should essentially be interconvertible with a meter state that is characterised by this parameter in the sense that $X(n)$ is roughly interconvertible with a $\tau_{n\lambda_{X_\infty}}$. The relation of equilibrium states to such macroscopic meter states concerns only their (approximate) behaviour, but they are not meter states (as opposed to~\cite{Lieb2014}).

\begin{defn} \label{def:equil}
$X_\infty$ is a \emph{thermodynamic equilibrium state} if there exists $\lambda_{X_\infty}$ such that for any $\delta > 0$ there exist meter states $\tau_{\lambda_\infty^{-}}, \tau_{\lambda_\infty^{+}} \in \Gamma_\mathrm{M}^{\infty}$  such that
\begin{equation}
\tau_{\lambda_\infty^{-}} \prec_\infty X_\infty \prec_\infty \tau_{\lambda_\infty^{+}} \ 
\end{equation}
and such that the parameters of the meter states $\lambda^{-}_n$ and $\lambda^{+}_n$ obey $\lambda^{-}_n \geq n(\lambda_{X_\infty}-\delta)$ and $\lambda^{+}_n \leq n(\lambda_{X_\infty}+\delta)$ for large enough $n$ respectively.
The set of all thermodynamic equilibrium states is denoted $\Gamma_\infty$.
\end{defn}

Thermodynamic equilibrium states are essentially characterised by a single entropic quantity, as shown with the following assertion that is based on Proposition~8.1.3 in~\cite{thesis}.

\begin{prop} 
\label{prop:aep}
Let there be an error-tolerant resource theory on state spaces $\Gamma(n)$ that obeys Axioms (A1) to (A4) and Axiom (A5) and let there be a suitable meter system on each of them. Furthermore, let $X_\infty \in \Gamma_\infty$ be a thermodynamic equilibrium state. 
Then, for any $0 < \eps < 1$,  
\begin{equation} \label{eq:aep}
S_{\infty}(X_\infty)\defeq\lim_{n \rightarrow \infty} \frac{S_-^{\eps}(X(n))}{n}= \lim_{n \rightarrow \infty} \frac{S_+^{\eps}(X(n))}{n}= \lambda_{X_\infty} \ .
\end{equation} 
\end{prop}

\begin{cor} \label{cor:aep}
Consider states $X \in \Gamma$ and macroscopic states $X_\infty \in \Gamma_\infty$, where  $X(n)=(X,\ldots,X)$ is the composition of $\ceil{n}$ copies of $X$ for each $n$. If all such states $X_\infty$ obey the requirements of Proposition~\ref{prop:aep}, i.e., if the $X(n)$ obey all axioms (including the consideration of suitable meter systems) and the $X_\infty$ are thermodynamic equilibrium states, then
\begin{equation}
S_-^0(X) \leq S_{\infty}(X_\infty) \leq S_+^0(X) \ .
\end{equation}
\end{cor}

The following shows how $S_\infty$ is the quantity that generally provides necessary and sufficient conditions for state transformations in the macroscopic regime. It is based on Proposition~8.1.5 of~\cite{thesis}.

\begin{prop} 
\label{lemma:typicality}
Let there be an error-tolerant resource theory on state spaces $\Gamma(n)$ that obeys Axioms (A1) to (A4) and Axiom (A5) and let there be a suitable meter system on each of them. Furthermore, let $X_\infty$, $Y_\infty \in \Gamma_\infty$ be thermodynamic equilibrium states. 
Then, for any $\delta > 0$, 
\begin{equation}
S_{\infty}(X_\infty) + \delta \leq S_{\infty}(Y_\infty) \implies X_\infty \prec_\infty Y_\infty \ .
\end{equation}
Furthermore, the converse holds for $\delta=0$. More generally, if for some $0 < \eps < 1$, there exists an $n_0$ such that for all $n \geq n_0$, the relation $X({n}) \prec^{\eps} Y({n})$ holds, then 
\begin{equation}
S_{\infty}(X_\infty) \leq S_{\infty}(Y_\infty) \ .
\end{equation}
\end{prop}

The proposition thus guarantees that $S_\infty$ is monotonic with respect to $\prec_\infty$ and that $S_{\infty}(X_\infty) < S_{\infty}(Y_\infty)$ implies $X_\infty \prec_\infty Y_\infty$.  
The macroscopic transformations according to $\prec_\infty$ are fully characterised by a necessary \emph{and} sufficient condition in terms of $S_\infty$ (except for pairs of states where $S_{\infty}(X_\infty) = S_{\infty}(Y_\infty)$).

The relation $\prec_\infty$ establishes the connection to the structure of traditional resource theories~\cite{Janzing2000_cost, Horodecki2003, Horodecki2003b, Horodecki2011, Renes2014} and (with appropriate composition and scaling operations) to the resource theories for macroscopic equilibrium thermodynamics~\cite{Lieb1998, Lieb1999, Lieb2001}. We will show this in the following. Let us define the composition of macroscopic states as
\begin{equation}
\Gamma_\infty \times \Gamma'_\infty \defeq \left\{  (X_\infty,~Y_\infty )  \ \middle| \  \begin{array}{lll} ( X_\infty,~Y_\infty ):& \mathbbm{R}_{\geq 0} &\rightarrow \left\{\Gamma(n) \times \Gamma'(n) \right\}  \\  &n  &\mapsto (X(n), Y(n)) \in \Gamma(n)\times \Gamma'(n) \end{array} \right\}.
\end{equation} 
The scaling with a parameter $\alpha \in \mathbb{R}_{\geq 0}$ leads to a state space $\alpha \Gamma_\infty$, obtained by scaling the parameter $n$. More precisely, a state $\tilde{X}_\infty = \alpha X_\infty \in \alpha \Gamma_\infty$ is obtained from $X_\infty \in \Gamma_\infty$ as the function that maps $n$ to $\tilde{X}(n)=X(\alpha n)$.
With these operations, $\prec_\infty$ obeys Axioms (E1)--(E4), which follows directly from the definitions. In addition, $S_\infty$ is extensive in the scaling and additive under  composition, as stipulated for a thermodynamic entropy function. To see this, let there be an equilibrium state $X_\infty$, meaning that for any $\delta >0$ and $\eps>0$ there is a $n_0$ such that for $n \geq n_0$,
\begin{equation}
\tau_{\lambda^-_{n}}\prec^\eps X(n) \prec^\eps \tau_{\lambda^+_{n}}
\end{equation}
with $\lambda^-_{n} \geq n (\lambda_{X_\infty}- \delta)$ and $\lambda^+_{n} \leq n (\lambda_{X_\infty}+ \delta)$. This implies that for $\alpha n \geq n_0$ also
\begin{equation}
\tau_{\lambda^-_{\alpha n}}\prec^\eps X({\alpha n}) \prec^\eps \tau_{\lambda^+_{\alpha n}},
\end{equation}
which in turn implies that $\alpha X_\infty$ is an equilibrium state with $S_\infty(\alpha X_\infty)=\lambda_{\alpha X_\infty}= \alpha \cdot \lambda_{X_\infty}$  (and similarly ${S_\infty((1-\alpha) X_\infty)}  = {\lambda_{(1-\alpha) X_\infty}}= {(1- \alpha) \cdot \lambda_{X_\infty}}$). The additivity under composition follows since the states $\tau_\lambda$ obey (M2).
From this it also follows that for $0 < \alpha < 1$, ${S_\infty(X_\infty)}={S_\infty((\alpha X_\infty, (1-\alpha) X_\infty ))}$.\footnote{Axiom (E5) is, however, not implied by this.}

\section{Emergence of macroscopicity in the statistical approach} \label{sec:examples}

To identify the thermodynamic equilibrium states with respect to adiabatic processes, let us first consider the set $\Gamma_\mathrm{M}$ defined in \eqref{eq:setT}, which is the set of all states $\rho \in \cS(\cH)$ that obey $\Hmin{\rho}=\Hzero{\rho}$ (see Proposition~\ref{prop:axioms_LYmicro}). Hence, $\Gamma_\mathrm{M}$ is the set of all states that obey $\maxEV \left( \rho \right)=\frac{1}{\rank (\rho)}$, where $\maxEV \left( \rho \right)$ denotes the maximal eigenvalue of $\rho$. These are the sharp states.

A class of thermodynamic equilibrium states with respect to the smooth adiabatic operations (and using the meter $\Gamma_\Lambda$) are the microcanonical states, 
\begin{equation}
\Gamma_\infty^{\mathrm{micro}}= \left\{ \rho_\infty \ \middle| \ \begin{array}{lll} \rho_\infty:& \mathbbm{R}_{\geq 0} &\rightarrow \left\{\Gamma(n) \right\}  \\  &n  &\mapsto \rho(n)=\frac{\Pi_{\Omega_\mathrm{micro}(nU,nV,n)}}{\Omega_\mathrm{micro}(nU,nV,n)} \end{array} \right\} \ ,
\end{equation}
where $\Omega_\mathrm{micro}$ is the microcanonical partition function and $\Pi_{\Omega_\mathrm{micro}}$ is the projector onto its subspace. 
For a microcanonical state $\rho_\infty \in \Gamma_\infty^{\mathrm{micro}}$,  $S_\infty$ is the entropy per particle known from statistical mechanics, $S_{\infty}(\rho_\infty)=\lambda_{\rho_\infty}= \log_2(\Omega_\mathrm{micro}(U,V,1))$.

In the following, we show that the so-called i.i.d.\ states are also a class of thermodynamic equilibrium states with respect to smooth adiabatic operations. The  proposition is based on Lemma~8.2.1 from~\cite{thesis}.

\begin{prop}
\label{lemma:typicalityNO}
For smooth adiabatic operations on quantum states with state space $\Gamma=\cS(\cH)$ and the meter system $\Gamma_\Lambda$,
\begin{equation}
\Gamma_\infty^{\mathrm{i.i.d.}}=\left\{ \rho_\infty  \ \middle| \  \begin{array}{lll} \rho_\infty:& \mathbbm{R}_{\geq 0} &\rightarrow \left\{\Gamma(n) \right\}  \\  &n  &\mapsto \rho(n)=\rho^{\otimes \ceil{n}} \end{array} \right\}
\end{equation}
is a set of equilibrium states. Furthermore, $\lambda_{\rho_\infty}= H(\rho)$ is the von Neumann entropy $H(\rho) \defeq -\tr{\rho \log_2 \rho}$.
\end{prop}

This agrees with previous results regarding the asymptotic equipartition property for i.i.d.\ states~\cite{TCR}, which recover the von Neumann entropy as the relevant quantity in the macroscopic regime.\footnote{In~\cite{TCR} the max entropy $H_0^{\eps}$ instead of $H_\mathrm{H}^{1-\eps}$ is considered, however, the relation has also been proven for the latter~\cite{Dupuis2013_DH}.} Note that Corollary~\ref{cor:aep} implies that thus $S_-^0(\rho) \leq S_+^0(\rho)$ for all $\rho \in \cS(\cH)$ (which we know to be true for $H_\mathrm{min}(\rho)$ and $H_0(\rho)$).

Notice that with respect to \emph{probabilistic} adiabatic transformations, the set of i.i.d.\ states, $\Gamma_\infty^\mathrm{i.i.d.}$, is not a set of thermodynamic equilibrium states according to Definition~\ref{def:equil}. To see this, let $\eps \leq \frac{1}{4}$, let $\delta=\frac{1}{100}$ and take a state $\rho=\frac{3}{4} \ketbra{0}{0} + \frac{1}{4} \ketbra{1}{1}$. Then $\rho^{\otimes n}$ has a maximal eigenvalue $\lambda_\mathrm{max}(\rho^{\otimes n})=\left(\frac{3}{4}\right)^{n}$ and its rank is $\rank(\rho^{\otimes n})=2^{n}$.
Now consider $\tau_{\lambda_n^{-}} \pmaj^\eps \rho^{\otimes n}$ and $\rho^{\otimes n} \pmaj^\eps \tau_{\lambda_n^{+}}$, i.e.,
\begin{align}
\tau_{\lambda_n^{-}} &\rightarrow (1-\eps) \rho^{\otimes n} + \eps \xi_1 \\
\rho^{\otimes n} &\rightarrow (1-\eps) \tau_{\lambda_n^{+}} + \eps \xi_2.
\end{align}
For there to be such transformations, the necessary conditions 
$\lambda_n^{-} \leq n \log_2{\left( \frac{4}{3} \right)} - \log_2{(1- \eps)}$ and $\lambda_n^{+} \geq n \log_2{(2)}$  have to be met,\footnote{This follows as the state $\tau_{\lambda_n^{-}}$ has to be chosen such that it majorises  $(1-\eps) \rho^{\otimes n}$. Furthermore, $\rho^{\otimes n}$ has to majorise $(1-\eps) \tau_{\lambda_n^{+}}$, hence its rank has to be smaller than that of $\tau_{\lambda_n^{+}}$.} 
which imply that $\lambda_n^{+} - \lambda_n^{-} \geq n \log_2{\left(\frac{3}{2} \right)} +\log_2{( 1- \eps)} > 2 n \delta$.

\section{Conclusion}
 
In this chapter, we have presented a unified axiomatic framework for thermodynamics. The phenomenological viewpoint taken to describe the transformations of thermodynamic systems in the axiomatic setting is complemented with microscopic models of these processes, leading to explicit entropic quantities that characterise state transformations. The connection between adiabatic processes according to Lieb and Yngvason and their analogue in the statistical approach is of conceptual interest, since it also connects entropy measures from information theory to their thermodynamic counterpart. It extends the well-known link between entropy in statistical physics and information theory~\cite{Jaynes1957, Jaynes1957_2} (which are both based on microscopic descriptions of systems) to  phenomenological thermodynamics.

Considering approximations and errors is necessary for the theoretical treatment of thermodynamics, even if these errors are so small that we usually do not notice them. The structure of our error-tolerant framework deviates from that of resource theories without this feature (including the latter as a zero-error case). The operationally significant quantities for characterising transformations under smooth adiabatic operations (the error-tolerant versions of adiabatic processes) are the smooth min and max entropies from the generalised entropy framework~\cite{Dupuis2013_DH} (cf.\ Proposition~\ref{prop:NO_axioms}). The emergence of macroscopic thermodynamics from our error-tolerant framework furthermore establishes the latter as a natural underlying structure. We recover the von Neumann entropy (for i.i.d.\ states) and the entropy per particle from statistical physics (for microcanonical states) as the quantities that specify the existence of state transformations in the macroscopic regime.

Our error-tolerant axiomatic framework applies to systems of any size, including microscopic systems. 
By introducing entropy meters without a continuous parameter we furthermore provide an axiomatic basis for describing the thermodynamics of microscopic systems without referring to large meter systems in order to specify their properties. In fact, the meter system could be of a size comparable to that of the (small) system of interest (which is usually accompanied with limitations in the precision).

Due to the axiomatic nature of our error-tolerant framework, it can also be applied to other resource theories. In particular, some preliminary work suggests that a resource theory of \emph{smooth thermal operations}~\cite{VanderMeer2017} also obeys our axioms. This implies that we can consider this resource theory in the same manner as the adiabatic processes and derive corresponding quantities $S_-^\eps$ and $S_+^\eps$, which will be quantities that generalise the free energy. Whether our framework is naturally applicable beyond the realm of thermodynamics, for instance to the resource theory of asymmetry~\cite{Gour2009}, remains an interesting open question.

So far, our axiomatic framework lacks the ability to describe situations where quantum side information about a system is accessible. It is known that this can lead to new types of phenomena such as a negative work cost for erasure~\cite{DelRio2011}. The quantities that characterise the existence of state transformations in such cases would likely  correspond to conditional entropies as known from information theory. An axiomatic framework that takes quantum side information into account would hence provide an axiomatic foundation for these entropies, paired with an operational meaning.

\subsection*{Acknowledgement}
Part of this research was made possible by the COST MP1209 network. MW is supported by the EPSRC (grant number EP/P016588/1). PhF acknowledges support from the Swiss National Science Foundation (SNSF) through the Early PostDoc.Mobility Fellowship No.\@ P2EZP2\_165239 hosted by the Institute for Quantum Information and Matter (IQIM) at Caltech, from the IQIM
which is a National Science Foundation (NSF) Physics Frontiers Center (NSF Grant
PHY-1733907), and from the Department of Energy Award DE-SC0018407.


\begingroup

\let\itshape\upshape

\bibliographystyle{unsrt}

\endgroup

\end{document}